\documentclass[RNAAS]{aastex62}
\usepackage{gensymb}

\begin{document}

\title{Inverse Chop Addition: Thermal IR background subtraction without Nodding}

\correspondingauthor{A.G.M. Pietrow}
\email{alex.pietrow@astro.su.se}

\author{A.G.M. Pietrow}
\affiliation{Institute for Solar Physics, Stockholm University, Albanova University Centre, SE-106 91 Stockholm, Sweden}

\author{L. Burtscher}
\affiliation{Leiden Observatory, Leiden University, P.O. Box 9513, 2300 RA Leiden, The Netherlands}

\author{B. Brandl}
\affiliation{Leiden Observatory, Leiden University, P.O. Box 9513, 2300 RA Leiden, The Netherlands}
\affiliation{Faculty of Aerospace Engineering, Technical University Delft, 2629 HS Delft, The Netherlands}


\begin{abstract}
    Due to the large size and mass of the secondary mirror on next generation extremely large telescopes it will not be possible to provide classical chopping and nodding as is used during mid-IR observations today. As a solution to this we propose an alternative approach to thermal background reduction called `inverse chop addition`. Here we use the symmetries of the thermal background to replace nodding, which allows us to get nearly identical background reductions while only using a special chopping pattern. The performance of this method was tested during technical time observations on VLT/VISIR. With this method, a higher observational efficiency can be obtained than with `classical chopping and nodding`, while achieving equally good reduction results. These results suggest that `inverse-chop addition` could be a good alternative for classical chopping and nodding on both current and next generation ground-based facilities.
\end{abstract}
\section{} 
\section{Introduction}
\begin{figure}
\begin{center}
\includegraphics[width=0.8\textwidth]{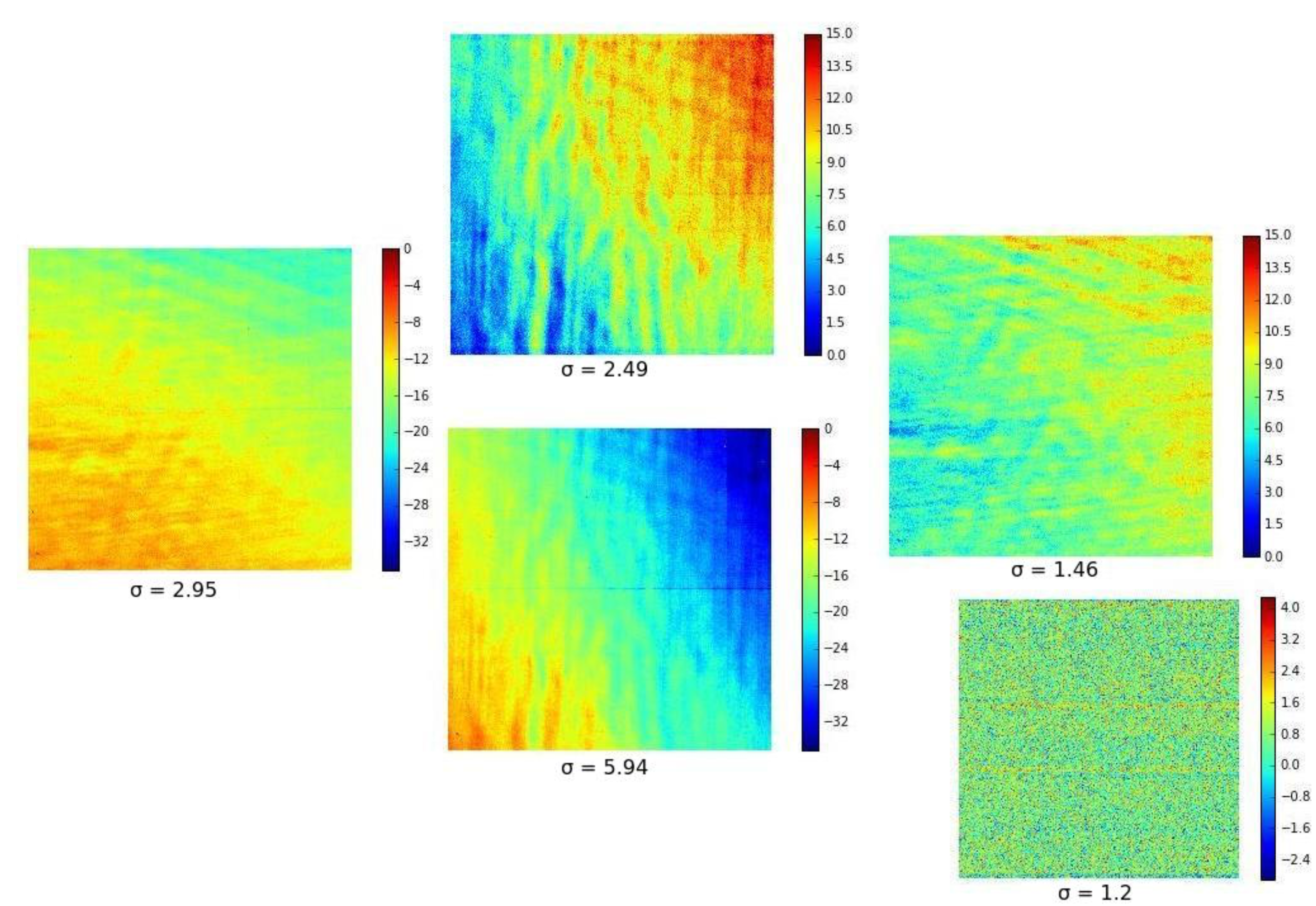}
\caption{Four panels illustrating the chop residuals obtained by chopping from the center to one of the four cardinal directions. The observations were obtained while tracking and de-rotation of the telescope were turned off, with the telescope pointing at zenith. 
We note that the structure of these background residuals is inverted in both structure and amplitude when compared to the residuals obtained from chopping ``down'' instead of ``up'' (or ``left'' instead of ``right''). In the lower right we see the resulting background that was obtained by adding the chop-up and chop-down frames together. An area void of sources was targeted during these observations. (Data taken on 22-03-2016.)} 
 \label{fig:2}
\end{center}
\end{figure}

Overall, the thermal background is the biggest challenge at infrared wavelengths beyond 2.5$\mu$m and results primarily from the warm telescope and atmosphere. At 12$\mu$m, this background is several orders of magnitude brighter than most astronomical sources. Very accurate background measurement and subtraction methods are required to reach the fundamental photon shot-noise limit of the thermal background. The classical method is the combination of chopping and nodding. In classical chopping, the secondary mirror is switching back and forth between the on-axis and an off-axis position. After subtracting the images obtained at these positions from each other, the background is greatly reduced. Since the thermal emission from the telescope structure is not uniform, and the on-axis and off-axis positions see slightly different parts of the optics, a ``chop residual'' (or ``thermal offset'') persists. That residual is usually removed by nodding, i.e. pointing the telescope to the off-axis position and repeating an inversed chopping sequence. In the ideal case, the residuals cancel each other and the observation is only limited by the shot noise of the background. In reality, the residuals do not vanish and the shot noise limit is often not reached.\\
\\
The Mid-infrared ELT Imager and Spectrograph (METIS) is one of three scientific first generation instruments on the European Extremely Large Telescope (ELT). METIS will cover the wavelength range between 3 - 19 $\mu$m  and will enable observations of dust-obscured and warm structures in the universe at five times the angular resolution of current 8 - 10m class telescopes. Due to the large size (4.2m) and mass of the ELT secondary mirror, it cannot provide classical chopping. While classical nodding is technically feasible, it would not be time efficient, due to overheads in the re-acquisition of the Adaptive Optics guide star. This necessitates novel approaches to thermal IR background removal.\\
\\\\
Several proposed methods could address this problem, including ``three beam chopping'' \citep{1992A&A...259..696L}, a weighted average method, similar to the LOCI algorithm \citep{2010SPIE.7736E..1JM}, ``drift scanning'' \citep{2014SPIE.9147E..9TH} and ``slow-scanning'' \citep{2018ApJ...857...37O}.  However, none of these methods have yet become mature and robust background subtraction technique for thermal-IR cameras at major observatories.\\
Therefore, we examined the origin and strength of chop residuals with VISIR on the VLT. We found that they seem to invert when chopping ``down'' instead of ``up'' (or ``left'' instead of ``right'') for not too large chop throws, as can be seen in Fig. \ref{fig:2}. \citep{pietrow2016} This implies that the chop residuals could be removed by chopping between three positions instead of two, where the two-off axis positions must be symmetric about the on-axis position. While this has already been shown to work with single-pixel detectors by \citet{1992A&A...259..696L}, it has never been demonstrated to work on array detectors, which usually require more accurate calibrations. In addition, our method differs from theirs due to limitations in the secondary mirror control of the VLT. Therefore, we employed an alternative approach, combining chop difference frames whose chop direction differs by 180\degree, a technique we call ``inverse chop addition''.\\
\\Both classical chopping and nodding and inverse-chop addition can be described with a simple mathematical equation. Using the notation as described in \citep{Vacca2018}, we find that the background for classical chopping and nodding can be defined as, 
\begin{equation}
n_A-n_B = \Delta B_{12} - \Delta B_{34} + 2S.
\end{equation} \textit{With $n_A$ and $n_B$ as the two nod positions, $\Delta B_{n,m}$ as the difference in background between the two chop positions of one nod position and S as the contribution by the source. Assuming that $\Delta B_{12}$ and $\Delta B_{34}$ have similar values, this leaves us with only the source.}\\\\
We can apply the same method method to inverse-chop addition and obtain,
\begin{equation}
c_A+c_B = \Delta B_{12} - \Delta B_{34} - \Delta O_{AB} + 2S.
\end{equation} \textit{With $c_A$ and $c_B$ as the same pointing with opposite chop directions, $\Delta B_{n,m}$ as the difference in background between the two chop positions of one pointing, $O_{AB}$ as the difference in the telescope path due to the movement of M2 and S as the contribution by the source.} \\ 
\\
This shows us that as long as the $\Delta O_{AB}$ term is small enough, the two methods should yield similar backgrounds after application. This term is thought to purely dependent on the size of the chop throw as can be seen in Fig. 3.4 of \citep{pietrow2016}.
\section{Observations}
The data for this research was taken at the VLT's VISIR instrument using several technical time slots during twilight. We first tested our method by observing a set of eight chop/nod observations with the telescope being "at rest". (tracking and de-rotation disabled) This was done in order to differentiate between effects of the sky and effects of the telescope. We observed through the J8.9 filter at a chop frequency of 4Hz, with a chop throw of 10'', and a position angle of 0, 90, 180, and 270 degrees. This allowed us to directly compare the results from classical chop/nod reductions with inverse chop addition under near identical conditions within the same dataset. (Certain permutations of these frames can be used to construct inverse chop addition pairs, while others can be used to obtain classical chopping and nodding). The observations were made close to zenith to test the method with fast field rotation. \\
Afterwards we repeated the same observation program, but this time under normal operating conditions, observing the star HR~2652. The latter observations were taken on the 24th of March, 2016, between 23:29:00 and 23:38:00 UT, again with the source being close to zenith to test the method with high on-sky rotation.\footnote{Data can be found in table 2.2 of \citep{pietrow2016} or by emailing the first author.}\\

\section{Results}
In order to compare classical chop nod reductions with inverse-chop addition we did a simple signal to noise measurement of HR~2652 after applying both methods, the results of this Can be seen in Fig. 2.\\
The first row of this figure shows the four directions of the chop throw in the data, with the corresponding nod frames in the second row. We note how the background is very similar in corresponding chop and nod pairs while it is inverted in opposing chop pairs. 
Adding the first two panels of the top row gives the first inverse-chop addition frame on the bottom row. Subtracting the first images of row one and two yields the classical chop/nod frame displayed as the second image of the bottom row. The same can be done to obtain the remaining two difference frames. We compare the effectiveness of the reduction using the mean of the background and the sum of the source. These are represented by the big red rectangle and the small black square, respectively. (See Table \ref{tab:1}.)\\
\\
For the vertically chopped frames we find a mean background signal of $0.07\pm1.20$ ADU for inverse chop addition, and $0.19\pm1.17$~ADU for classical chopping and nodding. For the horizontally chopped frames we find a mean background of $0.66\pm1.22$ ADU for inverse chop addition and $0.55\pm1.18$~ADU for classical chopping and nodding. The source counts are slightly lower in the inverse-chop added frames, giving us a S/N of 158,227 versus 162,574 for the vertical frames and 146,925 versus 156,096 for the (noisier) horizontal frames. This corresponds to a 0.4\% and 2.65\% decrease in signal respectively when comparing inverse-chop addition to classical chopping and nodding. The origin of this difference needs further investigation but might be caused by the difference term from Eq. 2.  Nevertheless, the high quality (low noise) of the inverse chop addition method looks very promising.\newpage
\section{Conclusions}
In conclusion, inverse chop addition is a useful alternative to classical chopping and nodding, giving us an identical background noise level at a much lower operational cost. It will be of particular relevance for thermal infrared observations at the ELT where nodding will only be available with significant operational overheads. We note that a precise knowledge of the chop residual structure is not required as long as it is symmetric about the chop origin. We have recently obtained calibration observations with VISIR in order to quantify the conditions and timescales under which the symmetry of the chop residual holds, which will form the basis of future work. Finally, unlike most other aforementioned methods, inverse-chop addition is compatible with current generation chopping mirrors.
\begin{deluxetable}{l|ll|ll}
\tablecaption{Data from bottom row of Fig. \ref{fig:1} comparing inverse-chop addition and classical chopping and nodding in two directions. The noise is given as the mean and standard deviation of the red rectangle in the figure. The source signal is given by the the sum of the counts in the black square of the figure. \label{tab:1}}
\tablehead{
& \multicolumn{2}{c}{{Vertical}} &  \multicolumn{2}{c}{{Horizontal}}
}
\startdata
 & Inverse-chop addition & Classical Chopping/Nodding & Inverse-chop addition & Classical Chopping/Nodding  \\\hline
 Noise & $0.07\pm1.20$  & $0.19\pm1.17$ & $0.66\pm1.22$  & $0.66\pm1.22$  \\
 Source & 189812 & 190560 & 179721 & 184620\\
 S/N & 158,227  & 162,574 & 146,925 & 156,096 \\
\enddata
\end{deluxetable}

\begin{figure}
\begin{center}
\includegraphics[width=\textwidth]{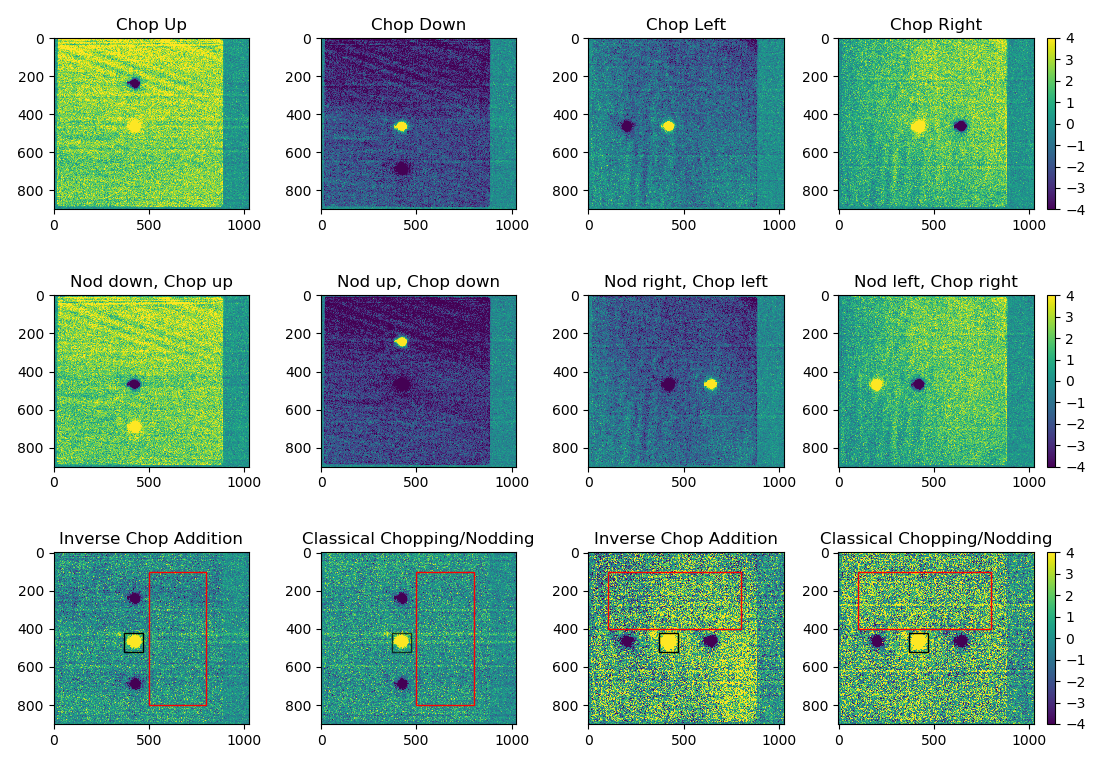}
\caption{The top row shows from left to right: the chop difference frames of HR2652 taken with chop angles of 0, 180, 270 and 90 degrees and the corresponding nod frames in the middle row. Note the inverted signal on the background pattern for opposing chop angles. The panels in the bottom row compare inverse chop addition with classical chopping and nodding along two chop directions. The red rectangle represents the region used for background calibration and the black square represents the region used for obtaining source counts. See table \ref{tab:1} for exact values.} 
 \label{fig:1}
\end{center}
\end{figure}

\subsection{Acknowledgments}
\noindent We thank Dr.~Konrad~Tristram (ESO~Chile) and Dr.~Daniel~Asmus (formerly ESO Chile, now at University of Southampton) for their help with obtaining our data and for fruitful discussions. \\ \\The Institute for Solar Physics is supported by a grant for research infrastructures of national importance from the Swedish Research Council (registration number 2017-00625).

\end{document}